\documentclass[3p,,preprint,12pt]{elsarticle}

\usepackage{lineno,hyperref}
\modulolinenumbers[1]
\usepackage{float}
\usepackage{amsmath}
\usepackage{amssymb}
\usepackage{pgfplots}
\usepackage{subcaption}
\usepackage{graphicx}

\usepackage{array}
\newcolumntype{P}[1]{>{\centering\arraybackslash}p{#1}}
 
\usepackage{setspace}
\usepackage{multirow}
\usepackage{color,soul}
\usepackage[figurename=Fig.]{caption}
\biboptions{sort&compress}
\AtBeginDocument{\hypersetup{pdfborder={0 0 0}}}
\begin{document}

\begin{frontmatter}

\title{\textbf{Delamination prediction in composite panels using unsupervised-feature learning methods with wavelet-enhanced guided wave representations}}

\author[a]{Mahindra Rautela \corref{contrib-0}}
\ead{mrautela@iisc.ac.in}\cortext[contrib-0]{Corresponding author}

\author[b]{J. Senthilnath}
\ead{j\_senthilnath@i2r.a-star.edu.sg}

\author[c]{Ernesto Monaco}
\ead{ermonaco@unina.it}

\author[a]{S. Gopalakrishnan}
\ead{krishnan@iisc.ac.in}

\address[a]{Department of Aerospace Engineering, Indian Institute of Science, Bangalore, India}
\address[b]{Institute for Infocomm Research, A*STAR, Singapore}
\address[c]{Department of Industrial Engineering, University of Naples Federico II, Italy}

\begin{abstract}
With the introduction of damage tolerance-based design philosophies, the demand for reliable and robust structural health monitoring (SHM) procedures for aerospace composite structures is increasing rapidly. The performance of supervised learning algorithms for SHM depends on the amount of labeled and balanced datasets. Apart from this, collecting datasets accommodating all possible damage scenarios is cumbersome, costly, and inaccessible for aerospace applications. In this paper, we have proposed two different unsupervised-feature learning approaches where the algorithms are trained only on the baseline scenarios to learn the distribution of baseline signals. The trained unsupervised feature learner is used for delamination prediction with an anomaly detection philosophy. In the first approach, we have combined dimensionality reduction techniques (principal component analysis and independent component analysis) with a one-class support vector machine. In another approach, we have utilized deep learning-based deep convolutional autoencoders (CAE). These state-of-the-art algorithms are applied on three different guided wave-based experimental datasets. The raw guided wave signals present in the datasets are converted into wavelet-enhanced higher-order representations for training unsupervised feature-learning algorithms. We have also compared different techniques, and it is seen that CAE generates better reconstructions with lower mean squared error and can provide higher accuracy on all the datasets.
\end{abstract}

\end{frontmatter}

\section{Introduction}\label{introduction}

Composite structures fail via different failure mechanisms, while delaminations are the most common ones which require proper attention. Delamination occurrence is considered a rare event in a newly commissioned aerospace structure. The number and size of these localized delaminations grow with time, accelerating towards the end life of the structure. Besides this, they are often hidden and undetectable by visual inspections. A safety-critical structural health monitoring (SHM) system demands anomaly detection at an earlier stage to avoid substantial structural residual strength reductions. This may consequently lead to a severe loss of life and infrastructure. Ultrasonic guided wave (UGW) based procedure is considered one of the popular techniques for SHM of aerospace structures. UGW is sensitive towards minor damages and can propagate long distances \cite{rautela2021temperature}. Due to these merits, UGW is used for damage identification in metallic and composite structures \cite{mitra2016guided}.

In recent years, supervised ML/DL techniques have shown promising results for guided wave-based damage detection \cite{liu2019deep,khan2019damage,zhang2020multi,rautela2021combined}. In most SHM applications like aerospace, collecting and storing big datasets is not feasible. It includes issues related to the labeling of the big dataset, which is prone to human errors. Apart from this, knowing about all possible damage scenarios in advance in varying environmental and operating conditions is a complex problem in itself. Therefore, the data collection process becomes cumbersome, time-consuming, and costly. It restricts the use of supervised ML/DL algorithms and encourages unsupervised ML/DL implementations. In most cases, manual features are generated and applied with supervised/unsupervised machine learning methods, which becomes tedious and requires domain expertise. Therefore, unsupervised-feature learning can benefit from discovering low-dimensional features that capture a meaningful representation of the high-dimensional input data.

Ruiz et al. \cite{ruiz2018multiway} have proposed dimensionality reduction methods like PCA with statistical control charts for guided wave-based unsupervised damage identification. In this method, PCA is used as a feature extractor by compressing the dimensionality of the input space. Relationship learners like SVM are used for unsupervised anomaly detection in building structures using wavelet transformed vibration signals \cite{he2007structural,ghiasi2016machine}. Different schemes are also presented in the literature to combine feature extractors and relationship learners. Zang et al. \cite{zang2004structural} have utilized a combination of ICA and artificial neural networks (ANN) for unsupervised damage detection in railway wheel and space antennas using vibrations signals. Khoa et al. \cite{khoa2014robust} have proposed PCA for feature extraction with one-class SVM (ocSVM) for anomaly detection in building-like structures. Wang \& Cha \cite{wang2021unsupervised} have combined autoencoders (feature extraction) with SVM (damage detection) on numerical and experimental bridge-like structures. From the aforementioned literature, it can be observed that both ML and DL-based feature-learning methods are equally preferred for anomaly detection in engineering structures. On the other hand, automatic feature extraction and relationship learning can be combined with deep autoencoders for anomaly detection. Silva et al. \cite{silva2021damage} have implemented a stacked autoencoders network for damage detection using two levels featurized vibration data. They have shown the implementations on Z-24 bridge structure.  Chow et al. \cite{chow2020anomaly} have utilized convolutional autoencoders for unsupervised visual inspection of concrete structures. 

Garcia et al. \cite{garcia2021temporal} have proposed different methods to convert one-dimensional time-series signals into images. They have performed transformation techniques like Gramian Angular Fields, Markov Transition field, recurrence plots, gray-scale encoding, spectrograms (based on Short-Time Fourier Transform), and scalograms (based on Discrete Wavelet Transform). It is observed that all encodings have improved the accuracy of anomaly detection, with scalograms providing the best results. Spectrograms and scalograms add an additional spectral representation in the one-dimensional temporal representation. Among these two popular time-frequency analysis schemes, scalograms provide better resolution in both time, and frequency domain for ultrasonic guided waves \cite{rautela2020ultrasonic}.

In our previous work, we have used convolutional autoencoders with wavelet-enhanced representations for delamination detection in composite panels \cite{rautela2021delamination}. This study is performed on a smaller scale, but the current study is much more extensive and detailed. More recent research by Wuttke et al. \cite{wuttke2021wave} have also used wavelet-transformation for propagating wavefields in a plate-like structure. A numerical dataset is collected from a dynamic lattice model, and an asymmetric encoder-decoder network is used to study crack initiation. Only a few investigations are conducted on unsupervised delamination detection problems in composite panels targeting aerospace-related applications in the literature. There is a limited emphasis made on the importance of higher-order representations for unsupervised feature learning methods. There is no evidence of the explicit comparative study of different unsupervised feature-learning methods.

In this paper, we have implemented both ML-based and DL-based feature learning methods for unsupervised delamination detection. In the ML-based approach, we have used PCA-ocSVM and ICA-ocSVM. Here, PCA and ICA are used for feature extraction using the linear dimensionality reduction, and ocSVM is used for relationship learning. In the DL-based approach, we have utilized sole convolutional-autoencoders to perform both feature extraction and relationship learning. We have used mean squared reconstruction error with different threshold values for unsupervised delamination detection. In both the approaches, we have trained the algorithms with baseline signals to learn the distribution of these signals and predict on the new baseline and damaged signals. We have compared both approaches in terms of reconstruction ability and testing accuracy. The algorithms are implemented on three different experimental datasets, i.e., Open Guided Waves (OGW) dataset \cite{moll2019open}, NASA Prognostic Center of Excellence-Guided Waves (NASA PCoE-GW) dataset \cite{peng2013novel}, and the University of Naples Computed Tomography Guided Waves (UoNCT-GW) dataset \cite{memmolo2015damage}. These datasets are different in terms of the geometry of the specimen used, material properties of the composites, layup type and sequence, number of sensors installed, sensor array geometry (parallel and circular), number of training and testing samples, excitation and sampling frequency, size and shape of damage. We have exploited continuous wavelet transformation (CWT) to transform guided-wave datasets containing time-series signals into higher-order time-frequency representations. The new wavelet-enhanced featurized space is used as input to the algorithms. The presented research work is novel in terms of (i) delamination prediction in composite panels using guided waves, (ii) performing different unsupervised feature learning methods for delamination detection, (iii) implementation and comparison of the algorithm on three different experimental guided wave datasets, (iv) utilization of wavelet-enhanced representations for better featurization of guided-wave experimental datasets, (v) detailed comparisons of unsupervised feature-learning methods.

The paper is organized as follows: Section-\ref{sec:background} contains the background of unsupervised feature learning methods, Section-\ref{sec:experiments} includes a brief introduction to the experimental setup \& procedures, Section-\ref{sec:training} presents training procedures and results, Section-\ref{sec:results} contains comparisons \& discussions. The paper is concluded in Section-\ref{sec:conclude}.
 
\section{Background: Unsupervised-feature learning methods} \label{sec:background}
\subsection{PCA-ocSVM \& ICA-ocSVM} \label{ssec:PCAocSVM}
PCA is one of the widely used linear dimensionality reduction technique. It works by orthogonal projection of data onto a lower dimensional principal subspace by maximizing the variance of the projected data. Mathematically, PCA can be posed as an unconstrained maximization problem using Lagrangian $L$ as shown in \cite{bishop2006pattern}:
\begin{gather}\label{eq:pca}
	\underset{u_{1}}{\mathrm{max}} \quad S_y =  u^T_{1}S_{x}u_{1} \nonumber\\
	\text{constraint: }u^{T}_1 u_{1} = 1 \\
	\underset{u_{1}}{\mathrm{max}} \quad L = u^T_{1}S_{x}u_{1} + \lambda_{1}(u^{T}_1 u_{1}-1) \nonumber
\end{gather}
\noindent Here, PCA is used to transform data ($x$) in D-dimensional vector space into 1-dimensional subspace ($y$) with unit vector $u_{1}$. $S_y$ is the variance of the projected data, $S_x$ is the data covariance matrix, and $\lambda_{1}$ is the Lagrange multiplier. It can be seen that $S_x$ is maximum when $u_1$ becomes its eigenvector with eigenvalue $\lambda_{1}$ \cite{ruiz2018multiway}. For a more general M-dimensional projection space, $\{u_1,u_2,...,u_M\}$ will be the eigenvectors of $S_{x}$ with eigenvalues $\{\lambda_{1},\lambda_{2},...,\lambda_{M}\}$. The loss of information due to the dimensionality reduction can be represented in terms of distortion measure, which is defined as the sum of eigenvalues orthogonal to the principal subspace. Explained variance ratio can also be utilized to quantify the loss of information. It is the ratio of the sum of eigenvalues of the principal subspace and the original vector space.

ICA is used to transform a vector space into a vector subspace having signals that are mutually independent. Statistically, it means a joint probability distribution of the independent signals is the product of probability distribution of each of them, i.e., $p(z_1,z_2,...z_n) = p(z_1)p(z_2)..p(z_n)$. In ICA, observed variables ($x$) are related linearly to the latent variables (independent components or source, s) but the latent distribution is non-Gaussian \cite{hyvarinen2000independent}. Mathematically, $x = As$, or $\hat{s} = Wx$, where, $x \in \mathbb{R}^d$ are the observed or measured variables, $\hat{s} \in \mathbb{R}^m$ are the reconstructed independent components, $A \in \mathbb{R}^{d\times m}$ is the mixing matrix and $W \in \mathbb{R}^{m\times d}$ is the demixing or separation matrix. The process of ICA includes centering, whitening and further processing steps. After centering $x$, it is linearly transformed into a new vector space $\tilde{x}$ where it becomes white (covariance matrix of $x$ is unity). One of the popularly used whitening method is eigenvalue decomposition i.e., $\tilde{x} = (ED^{-1/2}E^T)x$ where $E$ is a orthogonal matrix of eigenvectors of covariance matrix of $x$ and $D$ is the diagonal matrix having corresponding eigenvalues. This gives rise to a new orthogonal mixing matrix $\tilde{A} = ED^{-1/2}E^TA$ such that $\tilde{x} = \tilde{A}s$. The orthogonality of $\tilde{A}$ helps in reducing the number of parameters to be estimated from $n^2$ present in $A$ to $n(n-1)/2$ in $\tilde{A}$. After whitening, FastICA algorithms can be used to compute $\tilde{A}$, followed by calculating $\hat{s}$ \cite{hyvarinen2000independent}.

SVM is a widely used supervised learning technique implemented to map a feature space $x\in \mathbb{R}^d$ to labels, $y\in \{-1,+1\}$ with a hyperplane ($f:\mathbb{R}^d\rightarrow\{-1,+1\}$). The goal is to ensure minimum classification error on the dataset ($\{x_i,y_i\}^n_{i=1}$) along with maximum margin/width separating the two classes. SVM has a strong regularization property and provides a global optimum solution \cite{cortes1995support}. Mathematically, the minimization problem of SVM is defined as \cite{khoa2014robust}:
\begin{gather}
	\underset{w,\xi,b}{\mathrm{min}} \quad 0.5\|w\|^2 + C \sum_{i=1}^{n}\xi_i \nonumber\\
	\text{constraint: } y_i(w.x_i-b)\geq 1-\xi_i ,\quad \xi_{i} \geq 0 ,\quad i = 1,2,..,n
\end{gather}
\noindent Here, $w$ and $b$ are the training/learning parameters. The Euclidean norm of $w$ is defined by $\|w\|$, therefore, the first term of the equation is related to the distance between the hyperplane and the nearest data points of each class. $\xi_i$ controls training error and C balances $\xi_i$ (training error) and $w$ (the margin). The above problem can be converted into a dual form using Lagrange multipliers $\alpha_i$, $\alpha_j$:
\begin{gather}
	\underset{\alpha_1,\alpha_2,...,\alpha_n}{\mathrm{min}} \quad \sum_{i=1}^{n}\alpha_i - 0.5\sum_{i,j=1}^{n} \alpha_i \alpha_j y_i y_j x_i.x_j\\
	\text{constraint: } \sum_{i=1}^{n} \alpha_i y_i = 0 ,\quad 0\leq\alpha_i\leq C,\quad i,j=1,2,..,n
\end{gather}
The above problem can be solved using quadratic programming. For non-linear classification, kernel trick is used to replace the dot product by a non-linear kernel function, i.e., linear, polynomial, radial bias function and hyperbolic tangent \cite{khoa2014robust}. 

The one-class version of SVM called ocSVM can be used for anomaly detection. In unsupervised settings, the optimization problem can be formulated as \cite{scholkopf2001estimating}:
\begin{gather}
	\underset{w,\xi,b}{\mathrm{min}} \quad 0.5\|w\|^2 + \frac{1}{\nu n} \sum_{i=1}^{n}\xi_i - \rho \nonumber\\
	\text{constraint: } w.x_i \geq \rho-\xi_i ,\quad \xi_{i} \geq 0 ,\quad i = 1,2,..,n
\end{gather}

The hyperparameter $\nu$ in ocSVM plays a similar role as C of SVM. It controls the bias-variance tradeoff and consequently overfitting and generalization. The $\nu$ provides an upper bound on the training errors and a lower bound on the support vectors. If $\nu = 0.1$, it means 10\% of training samples are allowed to be misclassified. A higher $\nu$ gives a tighter decision boundary and more allowance for misclassifications. This is helpful in disregarding outliers during model construction. The goal of ocSVM is to create a hypersphere (decision boundary) around the baseline examples while maximizing the distance of this boundary from the origin. The classifier trained on baseline examples is used to classify test points lying outside the boundary as anomalies. 

\begin{figure}[h!]
	\centering
	\includegraphics[width=1.0\textwidth]{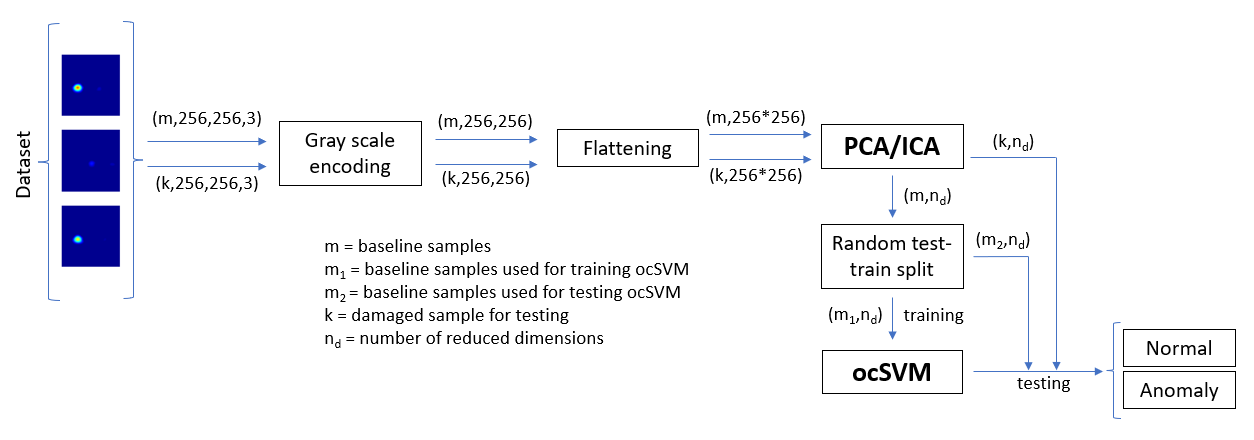}
	\caption{Detailed procedure for PCA-ocSVM \& ICA-ocSVM based delamination detection.}
	\label{fig:pcaicasvm}
\end{figure}

A detailed procedure for PCA-ocSVM and ICA-ocSVM based delamination detection is shown in Fig.~\ref{fig:pcaicasvm}. The process involves gray scale encoding and flattening the dataset and feeding into PCA/ICA for dimensionality reduction. The baseline dataset is randomly split into training ($m_1,n_d$) and testing datasets ($m_2,n_d$). The ocSVM is trained with baseline training dataset and tested for anomaly detection on baseline and damaged testing datasets. The accuracy is calculated and a confusion matrix is created based on the correctly classified and misclassified samples.

\subsection{Convolutional autoencoders} \label{ssec:CAE}
\begin{figure}[h!]
	\centering
	\includegraphics[width=0.45\textwidth]{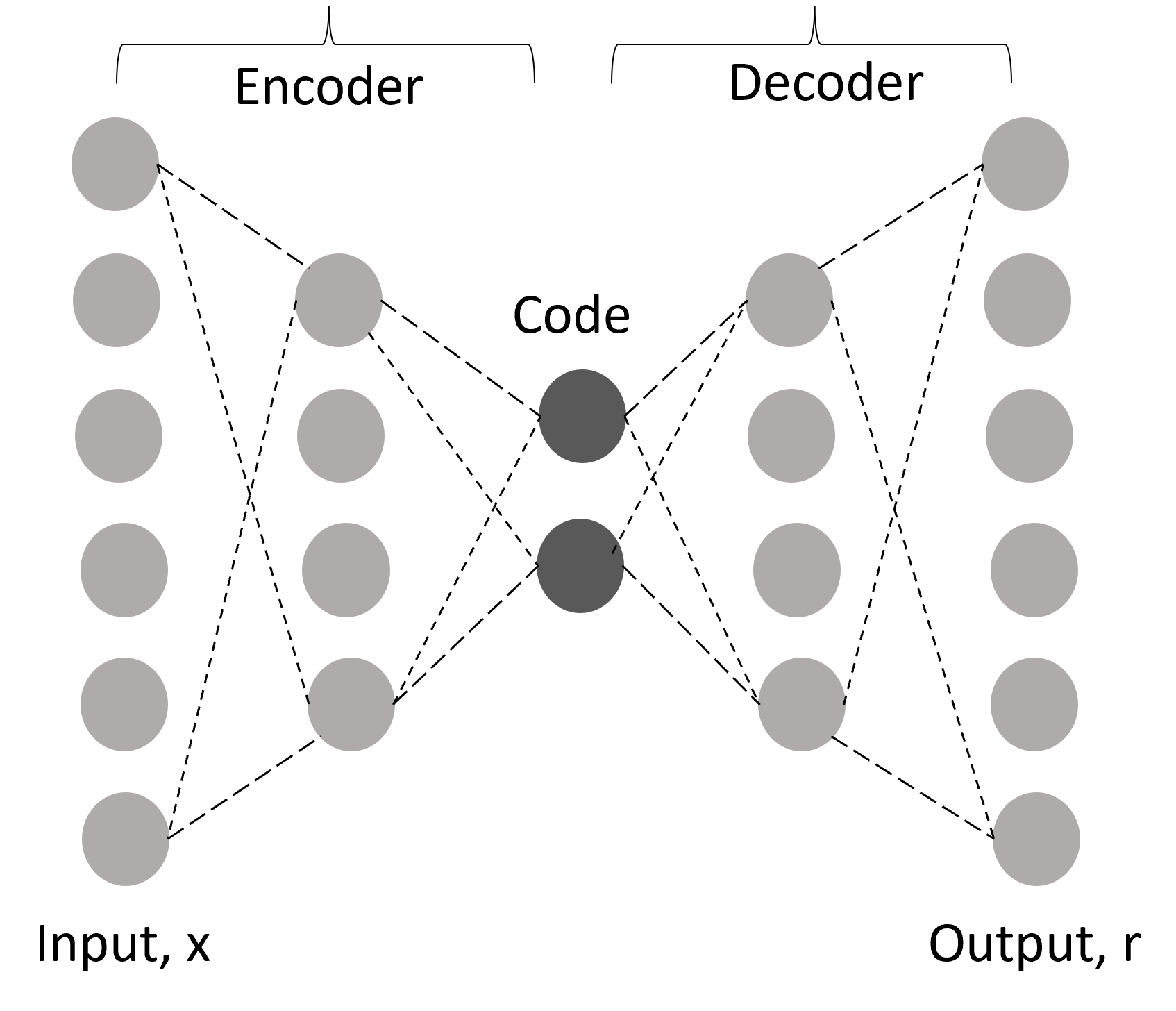}
	\caption{General architecture of a basic autoencoder. The data is compressed into the code layer and reconstructed back by the decoder.}
	\label{fig:ae}
\end{figure}

An autoencoder is a neural network-based technique that is trained to capture underlying distribution. It is different from supervised learning due to the absence of discrete (classification) or continuous (regression) labels, but input itself acts as a label. Therefore, autoencoders fall in the category of both unsupervised and self-supervised learning. It has three components, i.e., encoder, code, and decoder, as shown in Fig.~\ref{fig:ae}. An encoder compresses the input into code (or latent/hidden representation), and a decoder reconstructs back the input from the code layer. During this process, the code layer learns valuable features of the dataset. In a convolutional version of autoencoders, convolutional layers are used for encoding and decoding to incorporate parameter sharing and sparse connectivity into the network \cite{rautela2021inverse}.

Mathematically, an encoder is a deterministic mapping from d-dimensional input space to d'-dimensional latent space whereas a decoder maps from latent space to d-dimensional reconstructed input as shown \cite{vincent2008extracting}: .
\begin{equation}\label{eq:ae}
	\begin{split}
		h = f_{\theta}(Wx+b) \\
		r = g_{\theta'}(W'h+b')
	\end{split}
\end{equation}

where, $x$ is the input signals, $h$ is the latent space or hidden representation, $r$ is the reconstructed input or output, $\theta$ = [W,b] and $\theta'=[W',b']$ are the parameters of encoder and decoder, respectively. The parameters of encoder and decoder are optimized to minimize the reconstruction error, $L(x,g(f(x)))$ as shown:

\begin{equation}\label{eq:opt}
	[W,b,W',b'] = \underset{W,b,W',b'}{\mathrm{argmin}} \frac{1}{m} \bigg(\frac{1}{2} \Vert g(f(x^{i})) - x^{i} \Vert ^2\bigg)
\end{equation}

If $f_{\theta}$ and $g_{\theta'}$ are linear functions, $L(x,g(f(x)))$ is mean squared error and $d'<d$, autoencoders span the same subspace as PCA. With non-linear functions, autoencoders can learn more powerful generalization than PCA \cite{goodfellow2016deep}. 

\begin{figure}[h!]
	\centering
	\includegraphics[width=0.65\textwidth]{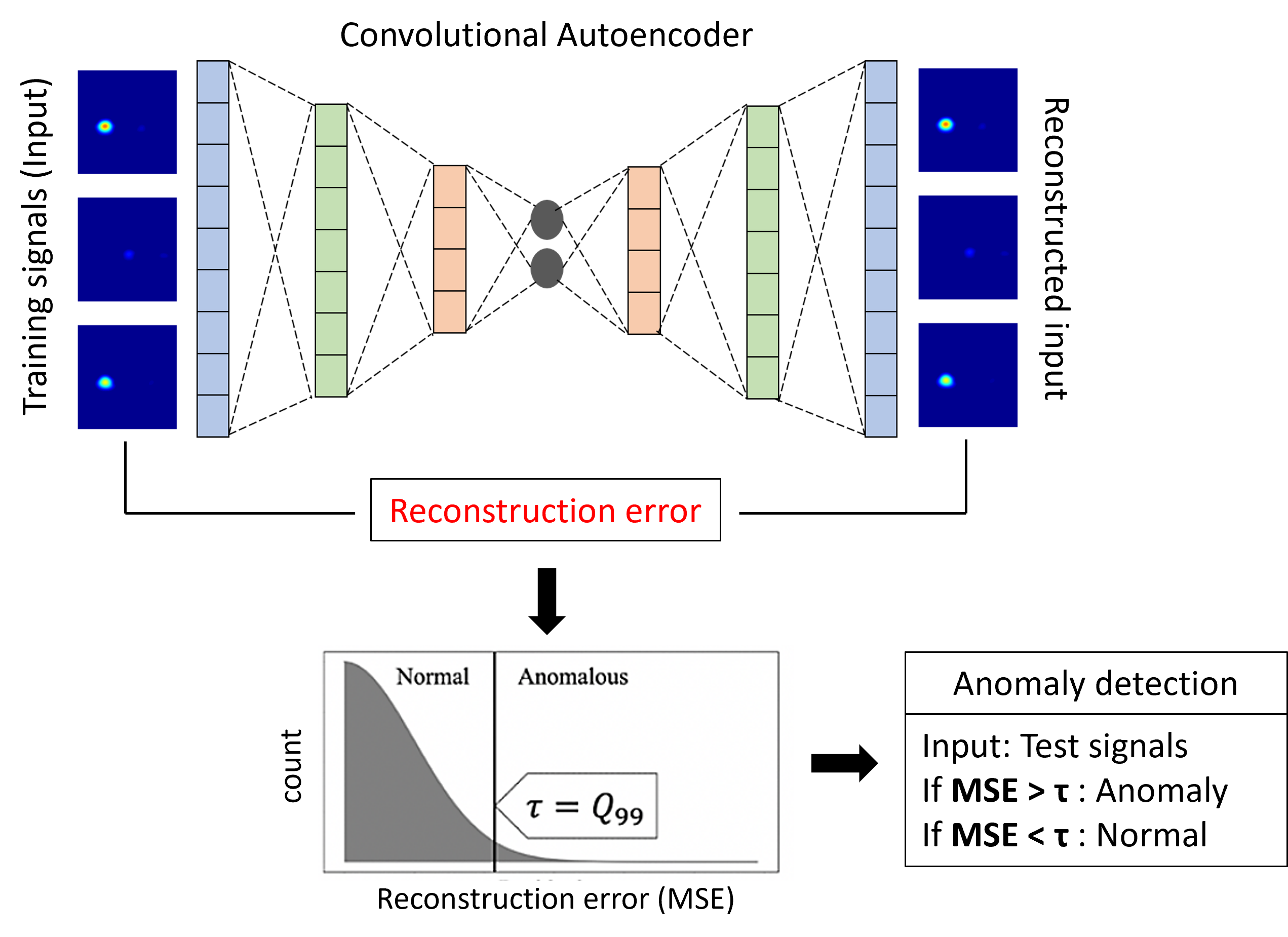}
	\caption{Convolutional autoencoders based delamination detection strategy.}
	\label{fig:strategy}
\end{figure}

In this work, we have used convolutional autoencoders for delamination identification. The strategy is illustrated in Fig.~\ref{fig:strategy}. In this procedure, CAE is trained with baseline signals to reconstruct its input. The mean squared reconstruction errors are collected for all the baseline samples. A threshold is decided to classify anomalous signals based on the distribution of error. The threshold settings depend on several factors and is application specific \cite{garcia2021temporal}. In this work, we have used maximum value and $99^{th}$ quartile of the reconstruction error. New signals are fed into the trained network and if the reconstruction error of the test set signal is more than the threshold, it is classified as an anomaly. If the error is less than the threshold, it is considered a normal signal.

\section{Experimental setup \& procedures} \label{sec:experiments}
\subsection{OGW dataset} \label{ssec:OGW}
\begin{figure}[h!]
	\centering
	\begin{minipage}[b]{0.48\textwidth}
		\centering
		\includegraphics[width=0.9\textwidth]{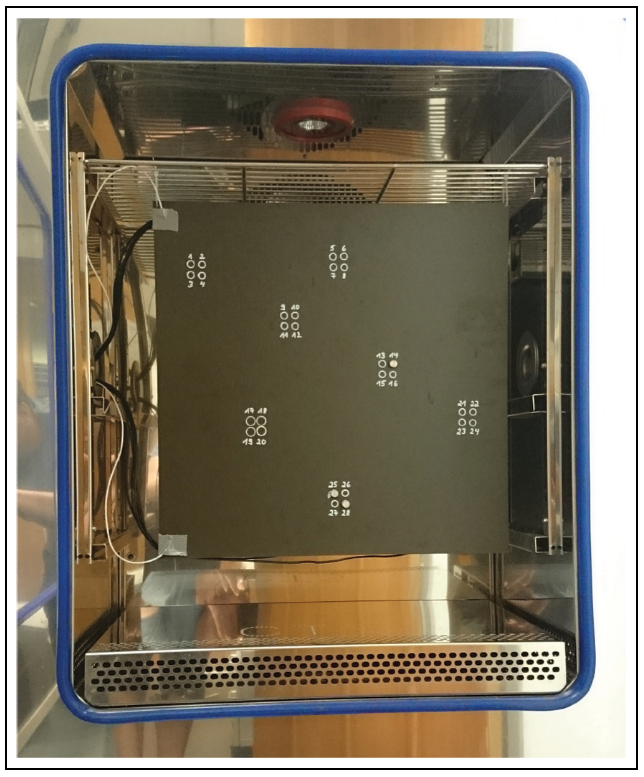}
	\end{minipage}
	\begin{minipage}[b]{0.48\textwidth}
		\centering
		\includegraphics[width=1.0\textwidth]{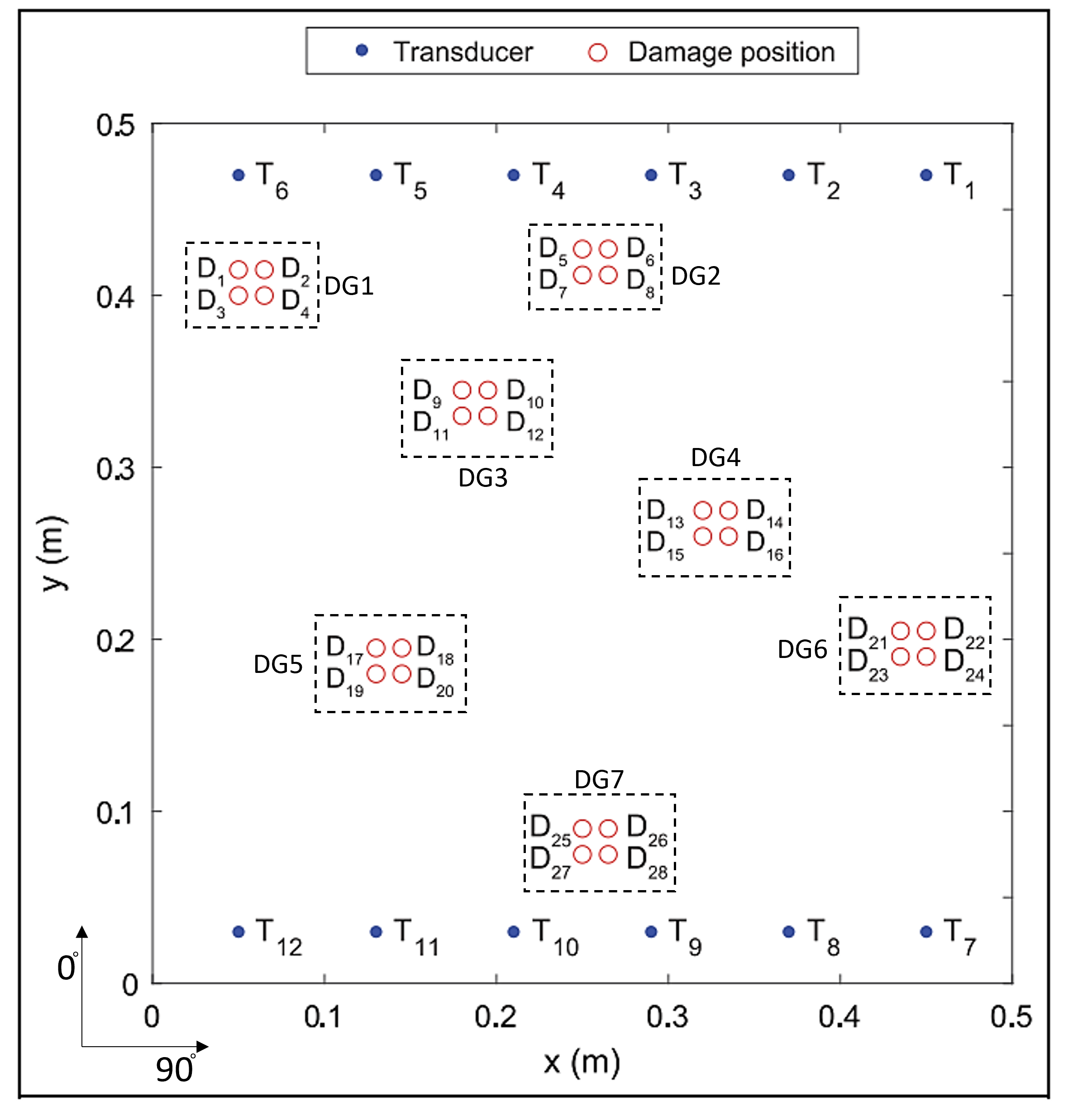}
	\end{minipage}
	\caption{Experimental setup (a) Panel placed in climate chamber at 23$^{\circ}$C and 50\% RH (b) Geometry of transducer position T1-T12 and defect locations D1-D28 \cite{moll2019open}.}
	\label{fig:ogw}
\end{figure}

The experimental setup along with the positions of transducers and damages is shown in Fig.~\ref{fig:ogw} \cite{moll2019open}. The experiments are performed on a 500$\times$500$\times$2 mm T700M21 carbon fiber reinforced polymer with a quasi-isotropic layup ([45/0/-45/90/-45/0/45/90]s), placed in a climatic chamber at 23$^{\circ}$C and 50\% RH. Twelve piezoelectric transducers are arranged in a parallel array giving rise to 66 signals per experiment when excited in a round-robin fashion with 5 cycles toneburst. Damages at 28 different positions are introduced using a reversible damage model in which an aluminum disk of 10 mm diameter and 3 mm thickness is bonded to the plate with double-sided adhesive tape. 

The experiments are conducted in six different phases. In the first phase, 20 baseline measurements were recorded from the healthy structure. In order to ensure fidelity and repeatability, multiple baseline measurements of the same experiment are performed. In the second phase, the reversible damage model is placed at 11 different positions, D1 to D11 and signals are recorded for every damage. Another 20 baseline measurements were recorded in the third phase, followed by measurements at damage positions D12 to D20 in phase 4. Similar procedure is followed in other next two phases. Twelve different frequencies ranging from 40 kHz to 260 kHz are used for excitation. A total of 47,520 baseline signals and 22,176 damage signals were collected from this procedure. The signals consist of fundamental symmetric ($S_0$) and antisymmetric ($A_0$) modes along with reflections from the boundaries \cite{rautela2021combined}. We are randomly picking 2500 baseline and 2500 damage signals for the training and testing purpose. However, we are running the algorithms ten different times to ensure repeatability of results.

\subsection{NASA-PCoE dataset} \label{ssec:NASA}

\begin{figure}[h!]
	\centering
	\includegraphics[width=0.7\textwidth]{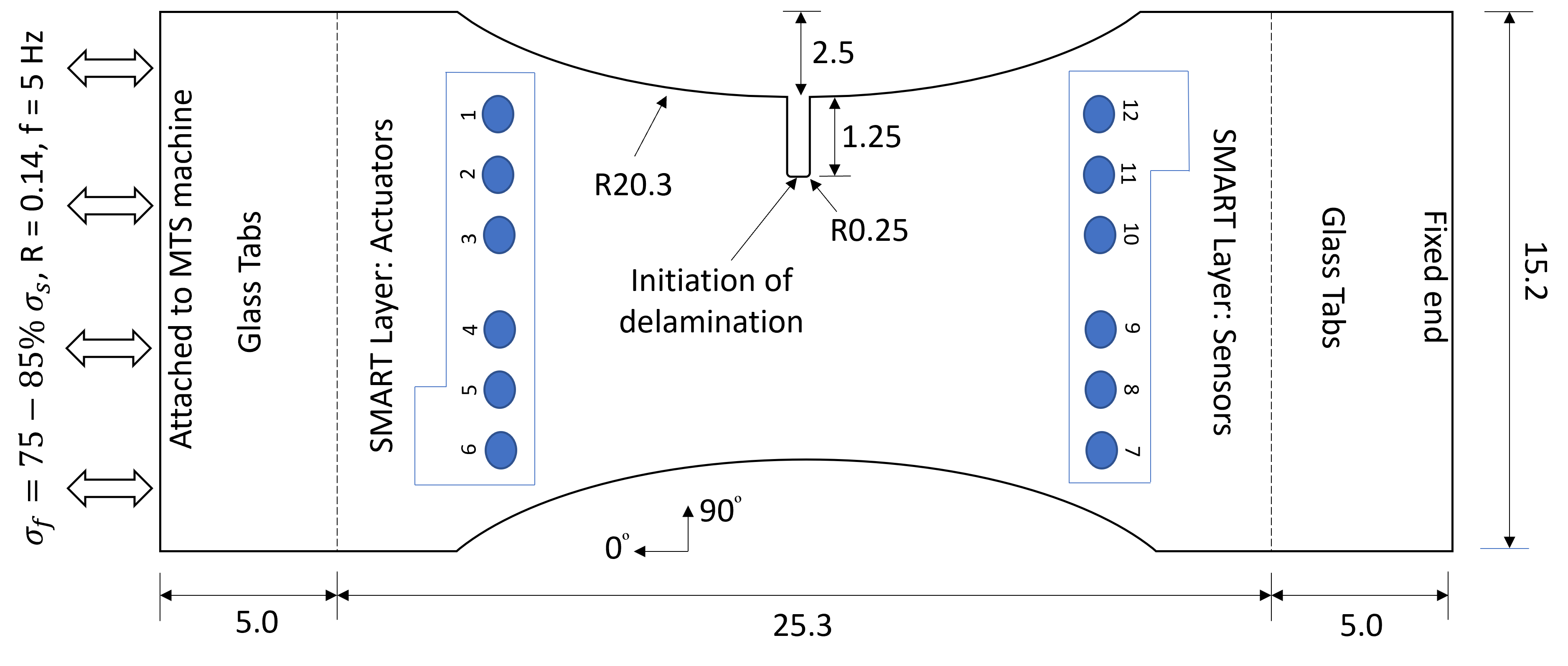}
	\caption{The geometry of dog-bone specimen used in NASA-PCoE dataset.}
	\label{fig:pcoe}
\end{figure}

For this dataset, fatigue experiments are conducted on a 15.24 cm $\times$ 15.24 cm dog bone specimen with an edge notch. The schematic representations is shown in Fig.~\ref{fig:pcoe} \cite{peng2013novel}. The specimen is made up of 12 plies uni-directional T700G composite material. Three symmetric layups are used from which we have considered quasi-isotropic layup-2 ([0/$90_2$/45/-45/90]s) for our work. Twelve PZTs are arranged in a parallel array on both sides of the specimen in which PZTs 1-6 are used for actuation, and PZTs 7-12 are used for sensing. Thirty-six actuator-sensor paths are recorded for each excitation frequency ranging from 150-450 kHz. Static run-to-failure tests are performed on an MTS machine with a load ratio of 0.14 following ASTM standards. The frequency of fatigue tests is 5 Hz with a sinusoidal profile. The fatigue tests are stopped after a certain number of cycles to record PZT sensor experiments. It is observed from X-ray radiography that the size of delaminations increases with the higher number of fatigue cycles. The objective of the entire experiment is to understand the effect of delaminations on the residual strength and durability of the component. In our study, we have used the dataset to predict the delamination once it has reached critical limit without investigating the complex problem of the time-dependent failure mechanism of the composite under cyclic loading. Here, the initial state of the structure is damaged with an edge notch. However, we can use this dataset considering the initial state as a baseline and upcoming states as delaminated \cite{wang2021integrating}.

In this study, we have used signals corresponding to 20k cycles. The total number of baseline and damages signals are 252 each. In order to increase the number of training examples to make the dataset suitable for deep learning applications, we have performed signal augmentation. In this work, we have added different small levels (SNR = 21.69 dB, 24.18 dB, and 27.69 dB) of random Gaussian noise into the 252 baseline training examples \cite{rautela2021combined}.

\subsection{UoNCT-GW dataset} \label{ssec:UoNCT}

\begin{figure}[h!]
	\centering
	\includegraphics[width=0.85\textwidth]{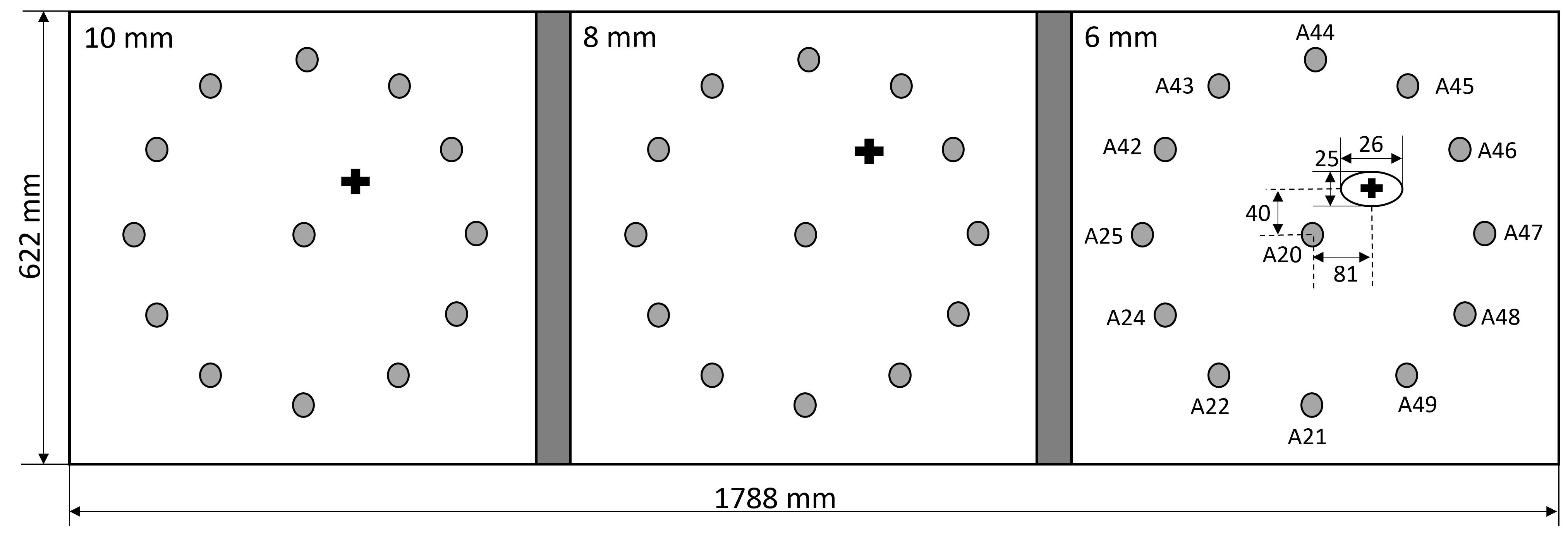}
	\caption{Tapered composite wing with three bays of different thickness. The impact position is represented by "+" sign. All dimensions in mm.}
	\label{fig:naples}
\end{figure}

In this experimental setup, a typical composite wing panel is used with three bays of different thicknesses (10 mm,  8 mm, and 6 mm) considering no wave interference between them \cite{memmolo2015damage}. Each bay is mounted with 13 PZTs in a circular array. The plate is laminated using the stacking sequence [5H/B45/U/B/U/B45/U/B]s where U are the unidirectional $0^{\circ}$ fibers, B are bidirectional laminae ($0^{\circ}/90^{\circ}$) fibers, B45 is $\pm45^{\circ}$ bidirectional laminae with top, and bottom layers of 5Harness (5H) plies \cite{memmolo2018damage}. In order to induce delamination, a drop weight with a 1-inch tip is used to impact three bays with a drop impact energy of 85, 110, and 150 Joules, respectively. The presence of delaminations is confirmed using C-scans. For this study, we have used the 6 mm bay as shown in Fig.~\ref{fig:naples}. The structure is excited with 4.5 cycles toneburst with a central frequency of 60 kHz and 80 V peak to peak amplitude in a round-robin fashion. The signals are recorded before and after the impacts, and the experiments are repeated ten times to ensure fidelity. A total of 1560 baseline signals and 1560 damage signals are collected in the entire process.

\subsection{Wavelet-enhanced guided waves representations}
\begin{figure}[h!]
	\centering
	\begin{minipage}[b]{0.32\textwidth}
		\includegraphics[width=1.0\textwidth]{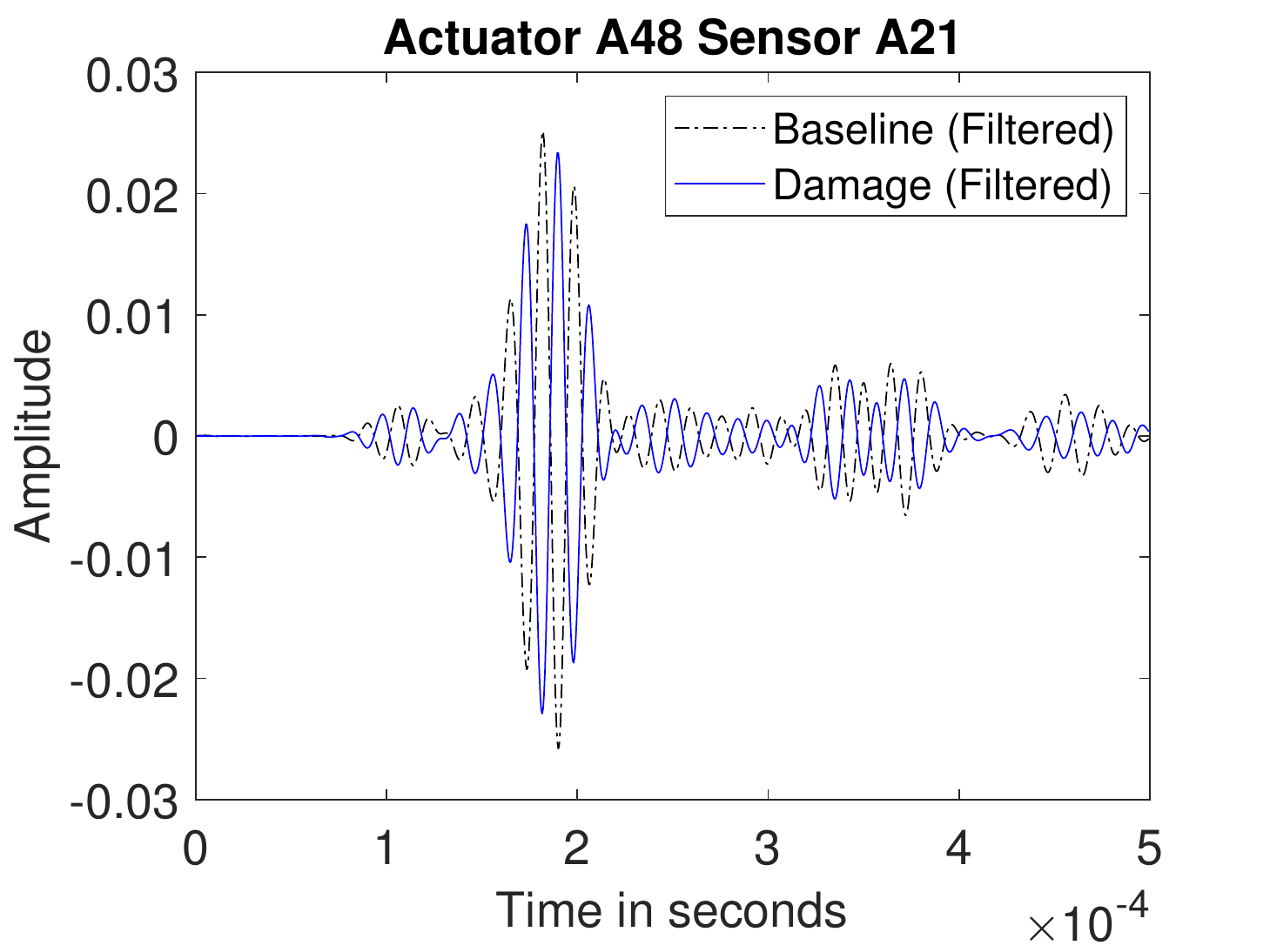}
	\end{minipage}
	\begin{minipage}[b]{0.32\textwidth}
		\includegraphics[width=1.0\textwidth]{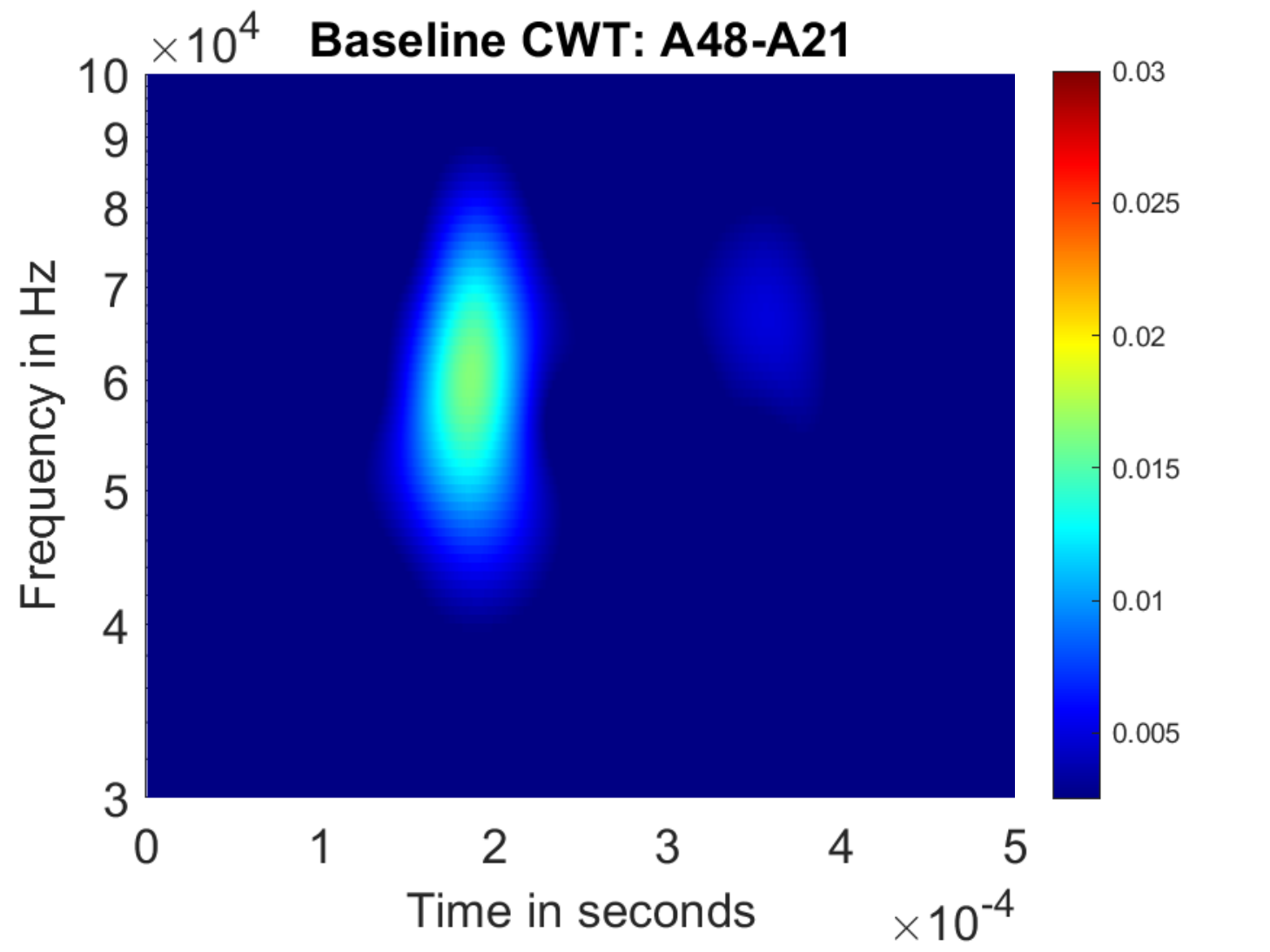}
	\end{minipage}
	\begin{minipage}[b]{0.32\textwidth}
		\includegraphics[width=1.0\textwidth]{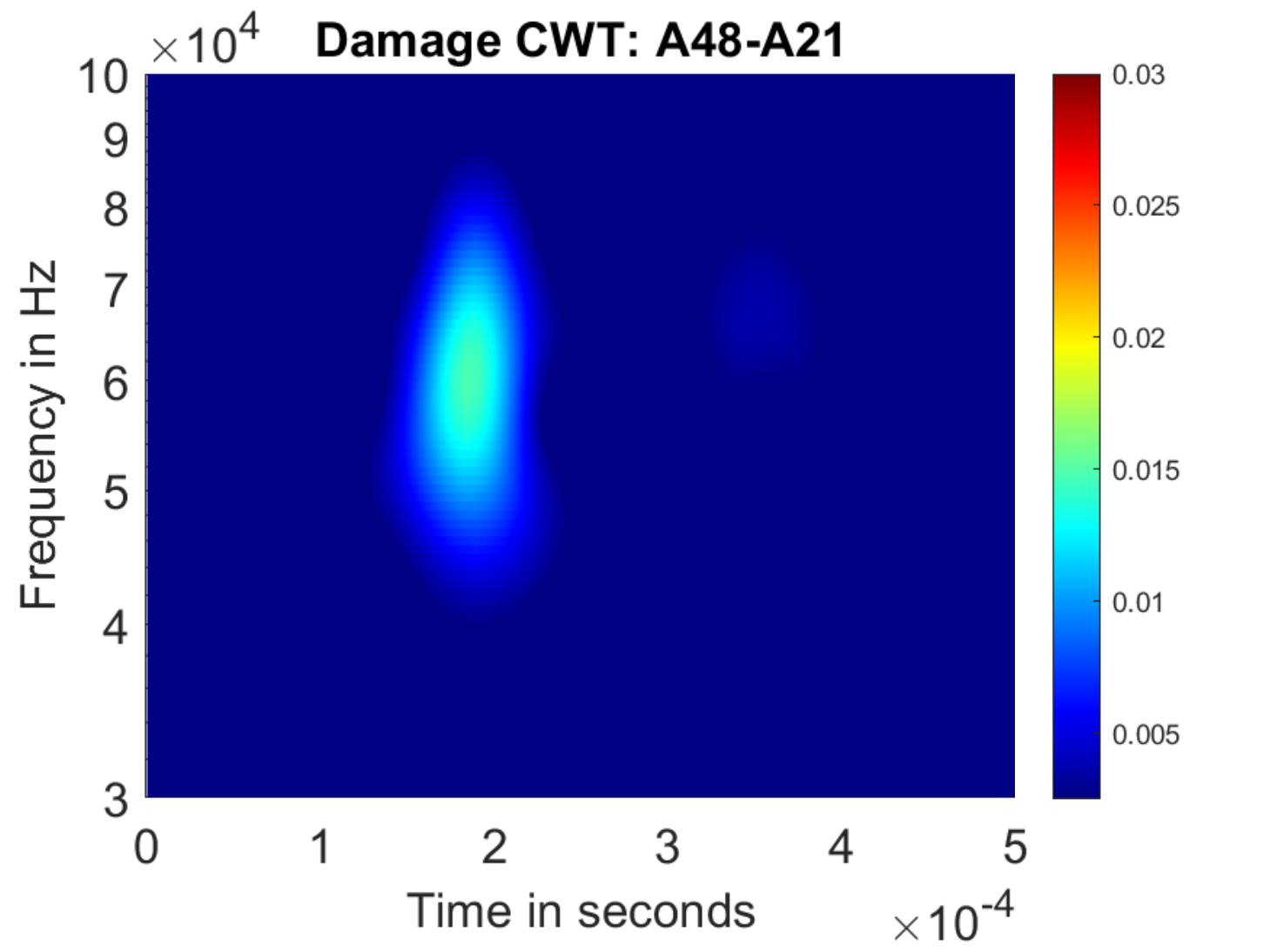}
	\end{minipage}
	\caption{Signals from UoNCT-GW dataset for actuator = A48 and sensor = A21 (a) Time-traces of baseline and damage signals(b) CWT representation of Baseline signal (c) CWT representation of damaged signal.}
	\label{fig:cwt}
\end{figure}

One-dimensional time-series signals can be converted into 2d time-frequency representations using time-frequency analysis. Short-time Fourier Transform (STFT) and Continuous Wavelet Transforms (CWT) are two popular techniques used for time-frequency analysis. It is reported that CWT provides better resolutions in time and frequency domains at higher frequencies \cite{abbate1995wavelet}. Since guided waves are high-frequency phenomena, CWT is considered suitable for representing time-series signals into time-frequency images. Autoencoders can be trained with time-series signals as well as with images. The reason for using time-frequency representation in the form of images is to provide networks with higher-order representations. This will help the network to extract better features and make training procedures easier \cite{garcia2021temporal}. Mathematically, CWT of a function $F(t)$ is represented as \cite{gopalakrishnan2010wavelet}.

\begin{equation}\label{eq:cwt}
	F^{W}(a,b) = \int_{-\infty}^{+\infty} F(t) \Phi \Bigg(\frac{t-b}{a}\Bigg)dt
\end{equation}

where, $\Phi$(t) is a wavelet basis function (analytical Morse wavelet). It is similar to hanning window in STFT. Here, `a' and `b' defines the width and position in time of $\Phi(t)$. They are used to modify $\Phi(t)$ by scaling and shifting it, which gives wavelet coefficients as a function of `a' and `b'. The wavelet coefficients are normalized and mapped on 256 different color intensities in the form of a heat map. Fig.~\ref{fig:cwt} shows a bandpass filtered time-series signal from the UoNCT-GW dataset and its corresponding CWT representations. Similar CWT is performed on all the time-traces of the three datasets.

In Fig.~\ref{fig:cwt}, the time-series plots show two dominant tonebursts, a more significant $A_0$ mode around 0.2ms and a smaller reflected $S_0$ mode near 3.5 ms. It is evident from the time traces that the damage signals have a lower amplitude (or energy) and are phase-shifted to the right than baseline signals. These characteristics are translated into time-frequency plots where (i) the bigger and smaller blobs are centered around the same timestamps, (ii) the ordinate axis represents the spread of spectral energy adding spectral representation, (iii) higher pixel intensity of baseline CWT plots corresponds to the higher energy of baseline signals, (iv) a smaller phase shift can also be visualized between both CWT representations.

\section{Training procedure and results} \label{sec:training}
\subsection{Training \& Testing datasets}
We have three datasets, and all of them are different in many aspects. We have compared the datasets based on criteria tabulated in Table~\ref{tab:datasets}. We have utilized time-frequency representations in the form of RGB images of shape 256$\times$256$\times$3 for training the algorithms. We have used 2125, 857, 1326 training examples and 2875, 403, 1794 testing examples for dataset-1, 2, and 3, respectively.

\begin{table}[h!]
	\caption{\small Datasets used for unsupervised feature-learning methods for delamination prediction.}
	\footnotesize
	\begin{center}
		\begin{tabular}{|l|c|c|c|}
			\hline
			Criteria & OGW Dataset & NASA Dataset & UoNCT Dataset \\
			\hline
			Specimen & Composite plate & Composite dog-bone & Composite panel \\
			Geometry & 500$\times$500$\times$2 & 25.3$\times$15.2 & 1788 $\times$ 622 $\times$ 6\\
			Lamina-type & Unidirectional & Uni. & Uni.,Bi.,5Harness \\
			Layup-type & Quasi-isotropic & same & same \\
			Layup Seq. & [45/0/-45/90/-45/0/45/90]s & [0/$90_2$/45/-45/90]s & [5H/$B_{45}$/U/B/U/$B_{45}$/U/B]s  \\
			Material & Carbon fiber-epoxy & same & same \\
			Sensor array & Parallel & Parallel & Circular \\
			Excitation & Round-robin & One-way & Round-robin \\
			Delamination & Model & Fatigue & Impact \\
			Damage shape & Circular & Semi-elliptical & Circular \\
			Damage size & 10 mm & 17 mm (major-axis) & 25 mm \\
			$N_{sensors}$ & 12 & 12 & 13 \\
			$F_{excitation}$ & 40-260 kHz & 150-450 kHz & 60 kHz \\
			$F_{sampling}$ & 10 MHz  & 1.2 MHz & 2 MHz\\
			$m_{train}$ & 2125 & 857 & 1326\\
			$m_{test}$ & 2875 & 403 & 1794 \\
			\hline
		\end{tabular}
	\end{center}
	\label{tab:datasets}
\end{table}

\subsection{ML approaches: PCA-ocSVM \& ICA-ocSVM}
\begin{figure}[h!]
	\centering
	\includegraphics[width=1.0\textwidth]{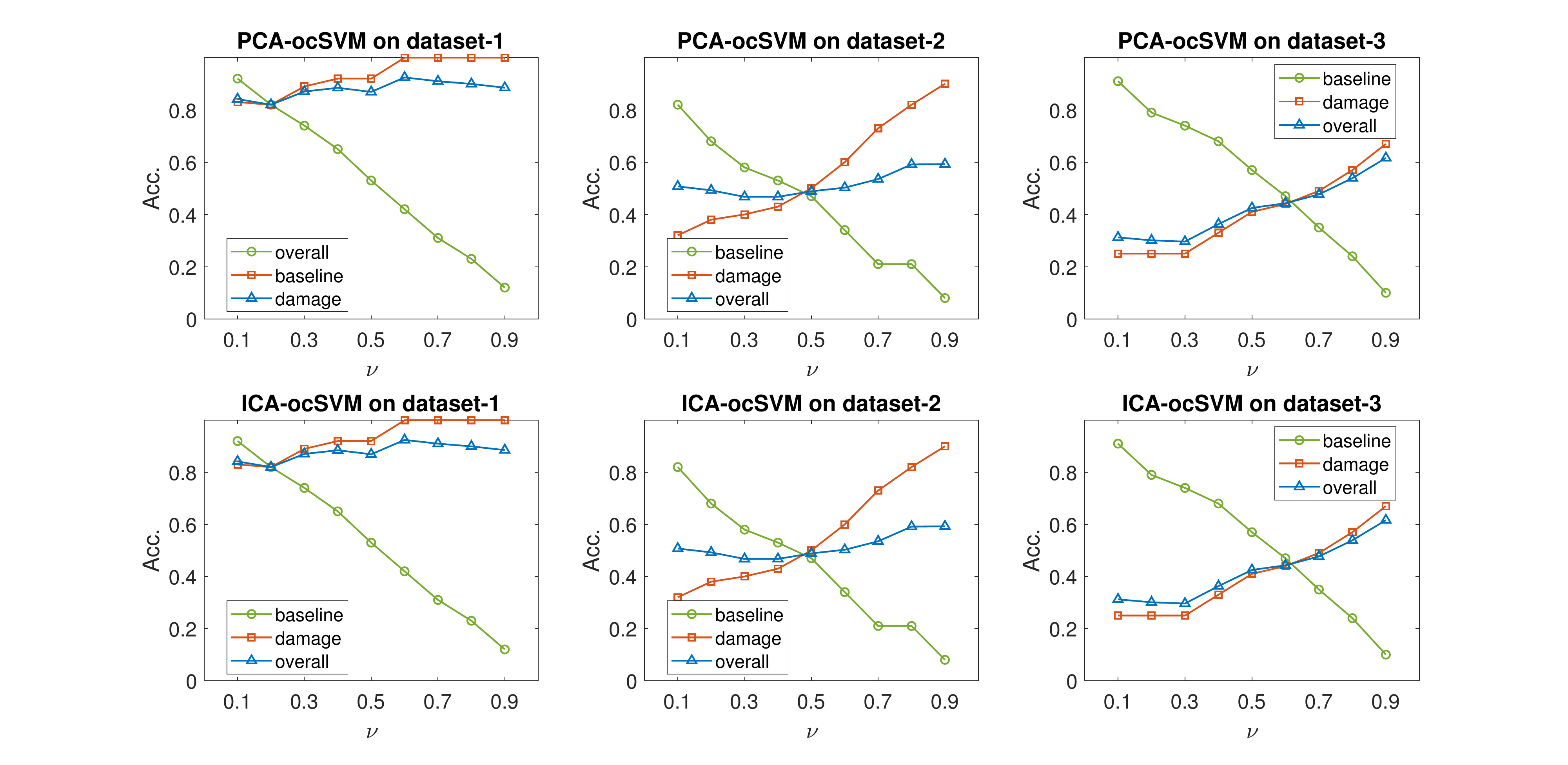}
	\caption{Testing accuracy vs hyperparameter $\nu$ for PCA-ocSVM (first Row) and ICA-ocSVM (second Row) across all 3 datasets.}
	\label{fig:accml}
\end{figure}

The training procedure for PCA-ocSVM \& ICA-ocSVM is shown in Fig.~\ref{fig:pcaicasvm}. For both of the techniques, we have reduced the input space to three-dimensional subspace. The selection of the reduced dimension is based on factors such as maximum explained variance and minimum mean squared reconstruction error. For PCA, we have used full singular value decomposition. For ICA, we have implemented FastICA algorithm with eigen value decomposition for whitening and a tolerance of 1e-4 and 500 iterations.  The time-taken by PCA and ICA are of order of few minutes on all the datasets. The PCA/ICA reduced baseline dataset is used to train ocSVM with the radial bias function (RBF). The accuracy is presented as function of hyperparameter $\nu$ across the three datasets in Fig.~\ref{fig:accml}. The value of $\nu$ is varied between 0 and 1 with an increment of 0.1.

For low $\nu$ like 0.1, true positives are given more weightage over false negatives whereas for higher $\nu$ like 0.9, true negatives are preferred over false positives. It can be seen that $\nu$ in ocSVM also controls the safety-cost tradeoff. Lower $\nu$ reduces false alarms, which in turn bring down inspections costs on the account of compromise on safety. On the other hand, higher $\nu$ increases the rate of false alarms and consequently inspection costs. However, this characteristics plays an important role while designing safety-critical systems. 

\subsection{DL approach: Convolutional autoencoders}
In convolutional autoencoders (CAE), an autoencoder is implemented with convolutional layers on all the three datasets. We have used mean squared error (MSE) based reconstruction loss as the cost function. We have utilized mean absolute error (MAE) and coefficient of determination ($R^2$) as performance metrics. We have followed the Adam optimization scheme to incorporate momentum and adaptive learning rate in the training process \cite{kingma2014adam}. We have used a learning rate of 1e-3 and a batch-size of 64, 32 and 20 for Dataset-1, 2 and 3 respectively, considering the smoothness of the loss curve and the generalization in the testing phase. The architecture of the autoencoder for delamination prediction is shown in Table.~\ref{tab:AEmodel}. There are three parts in the architecture i.e, an encoder (Conv2D$\rightarrow$BN$\rightarrow$....$\rightarrow$Conv2D$\rightarrow$BN$\rightarrow$Dense), hidden representation (Dense Layer with 3 neurons) and a decoder ($\text{Conv2D}^T$$\rightarrow$BN$\rightarrow$...$\rightarrow$$\text{Conv2D}^T$). 

\begin{table}[h!]
	\centering
	\footnotesize	
	\caption{\small Architecture of CAE for delamination prediction.}
	\addtolength{\tabcolsep}{-1pt} 
	\begin{tabular}{l|c|c}
		\hline
		Layer & Output Shape & Parameters\\
		\hline
		\textbf{Encoder}: &&\\
		Input Layer & (256, 256, 3) & 0 \\
		Conv2D (LReLU, 3x3, str = 2, 16) & (128, 128, 16) & 448 \\
		BatchNorm & (128, 128, 16) & 64\\
		Conv2D (LReLU, 3x3, str = 2, 32) & (64, 64, 32) & 4640 \\
		BatchNorm & (64, 64, 32) & 128\\
		Conv2D (LReLU, 3x3, str = 2, 64) & (32, 32, 64) & 18,496 \\
		BatchNorm & (32, 32, 64) & 256\\
		Conv2D (LReLU, 3x3, str = 2, 128) & (16, 16, 128) & 73,856 \\
		BatchNorm & (16, 16, 128) & 512\\
		Conv2D (LReLU, 3x3, str = 2, 256) & (8, 8, 256) & 295,168 \\
		BatchNorm & (8, 8, 256) & 1024\\
		\hline
		\textbf{Code}: &&\\
		Flatten & (16,384) &  0 \\
		Dense(50) & (50) & 819,250\\
		Dense(3) & (3) & 153\\
		& (16,384) & 65,536\\
		\hline
		\textbf{Decoder}: &&\\
		Mirrors the architecture of the Encoder&&\\
		\hline
		\hline
		Total parameters for Encoder &-& 1,213,995 \\
		Total parameters for Decoder &-& 1,049,955 \\
		\hline
	\end{tabular}
	\label{tab:AEmodel}
\end{table}

\begin{figure}[h!]
	\centering
	\begin{minipage}[b]{1.0\textwidth}
		\centering
		\includegraphics[width=0.9\textwidth]{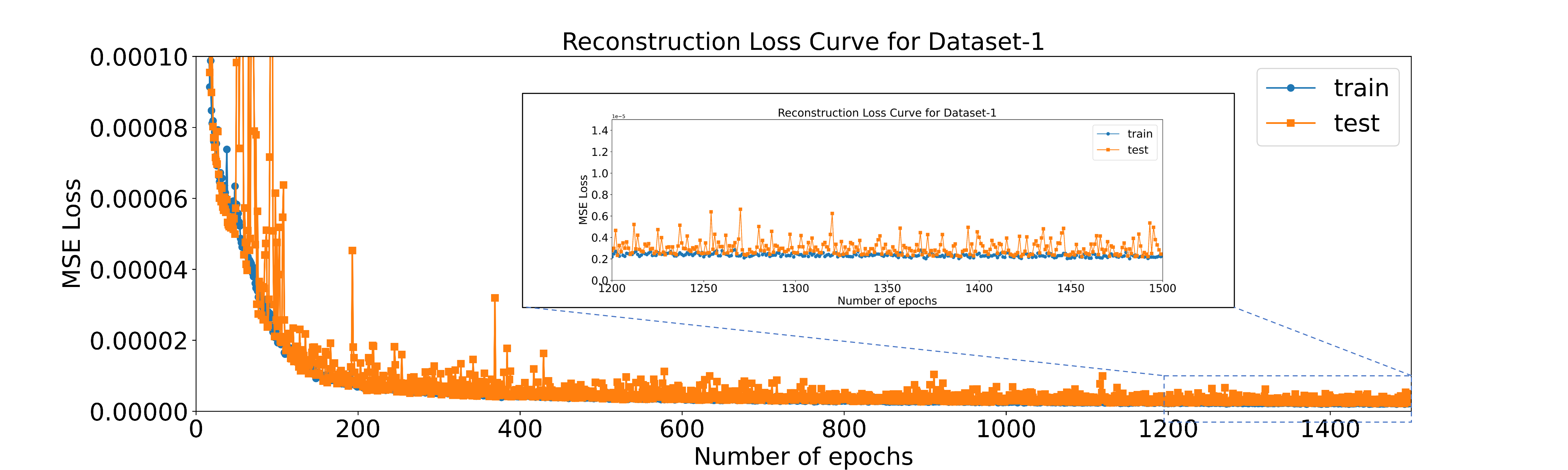}
	\end{minipage}
	\begin{minipage}[b]{1.0\textwidth}
		\centering
		\includegraphics[width=0.9\textwidth]{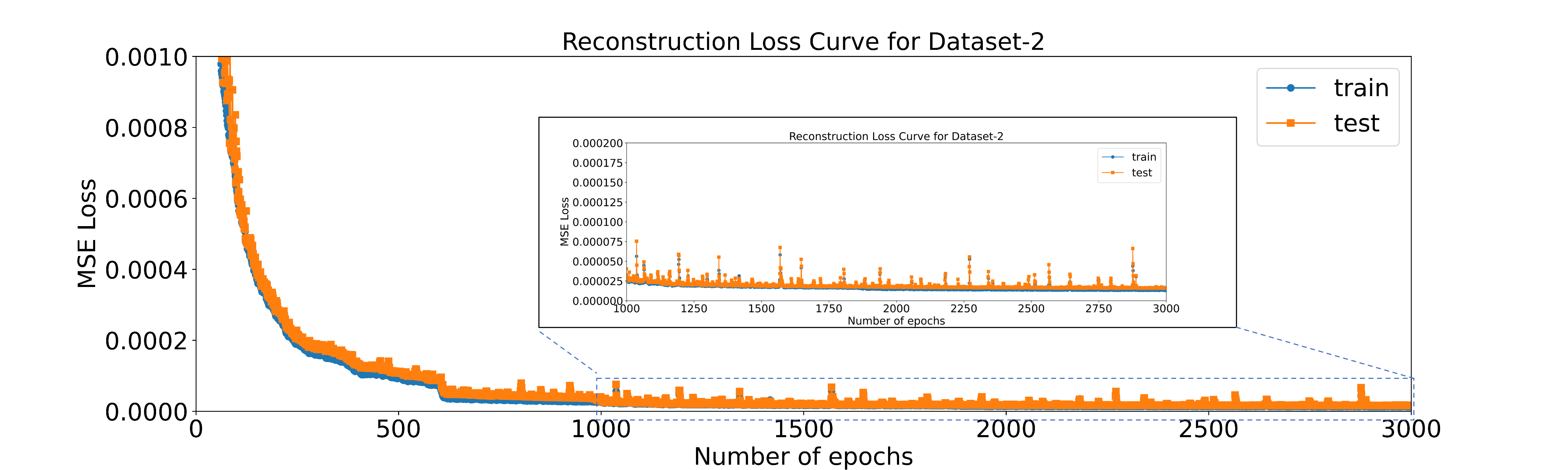}
	\end{minipage}
	\begin{minipage}[b]{1.0\textwidth}
		\centering
		\includegraphics[width=0.9\textwidth]{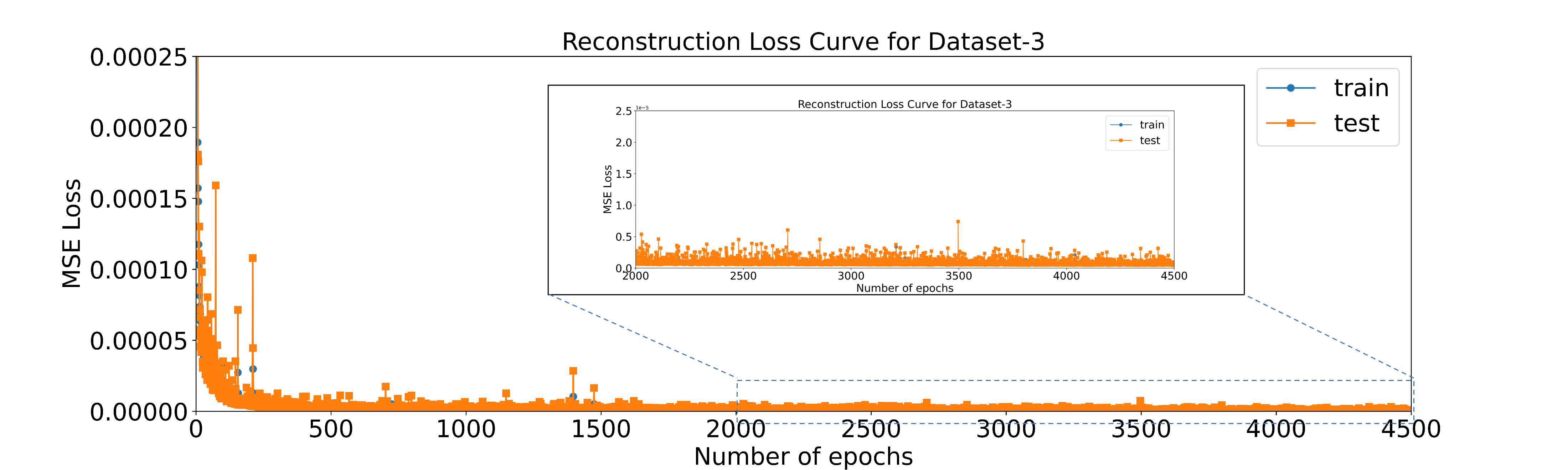}
	\end{minipage}
	\caption{Mean squared reconstruction loss for CAE based delamination prediction on (a) Dataset-1, (b) Dataset-2, and (c) Dataset-3.}
	\label{fig:loss}
\end{figure}

A kernel size of 3$\times$3 with filters: 16, 32, 64, 128, 256 with a stride length of 2 and a Leaky ReLU activation function are used in different convolutional layers. The total number of training parameters of the network is 2.2M. The MSE reconstruction loss is shown in Fig.~\ref{fig:loss}. The networks are trained for 1500, 3500, and 1500 epochs, taking 14, 5, and 6 seconds for each epoch, respectively. We have obtained average training reconstruction loss of 2.5e-7, 5e-7 and 8e-7 on dataset-1, 2 and 3 respectively.

\begin{figure}[h!]
	\centering
	\begin{minipage}[b]{1.0\textwidth}
		\centering
		\includegraphics[width=0.9\textwidth]{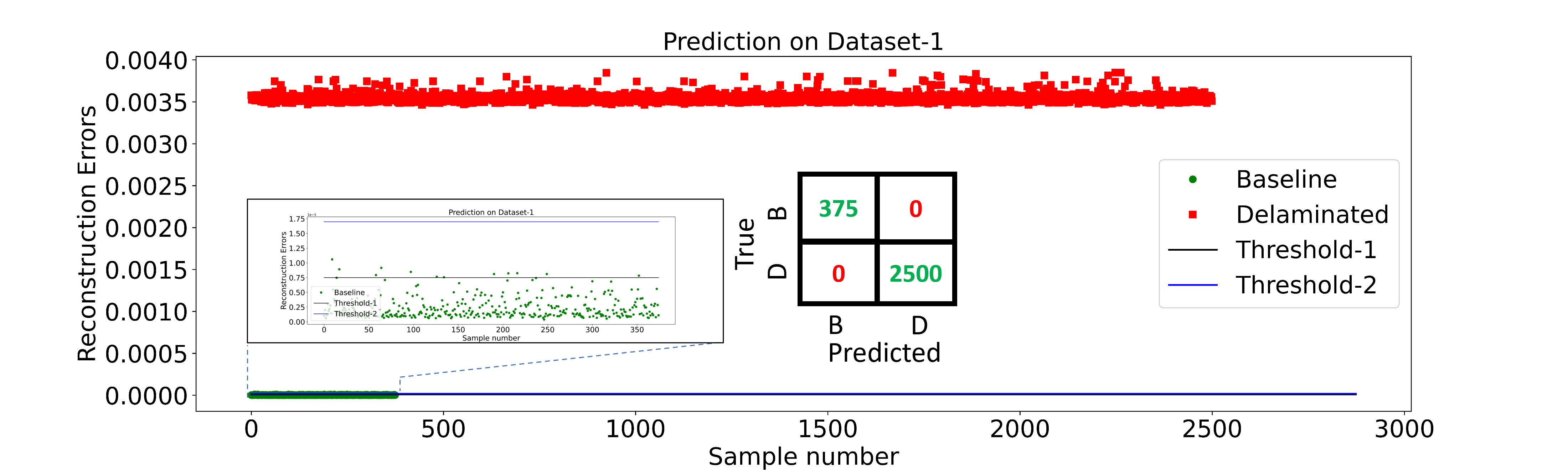}
	\end{minipage}
	\begin{minipage}[b]{1.0\textwidth}
		\centering
		\includegraphics[width=0.9\textwidth]{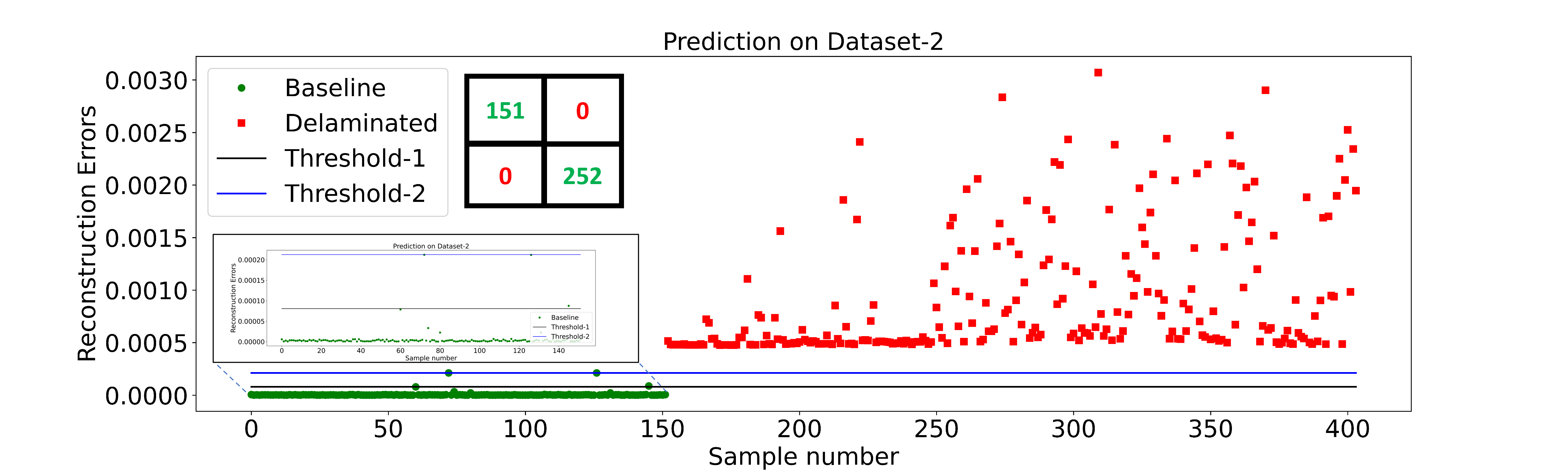}
	\end{minipage}
	\begin{minipage}[b]{1.0\textwidth}
		\centering
		\includegraphics[width=0.9\textwidth]{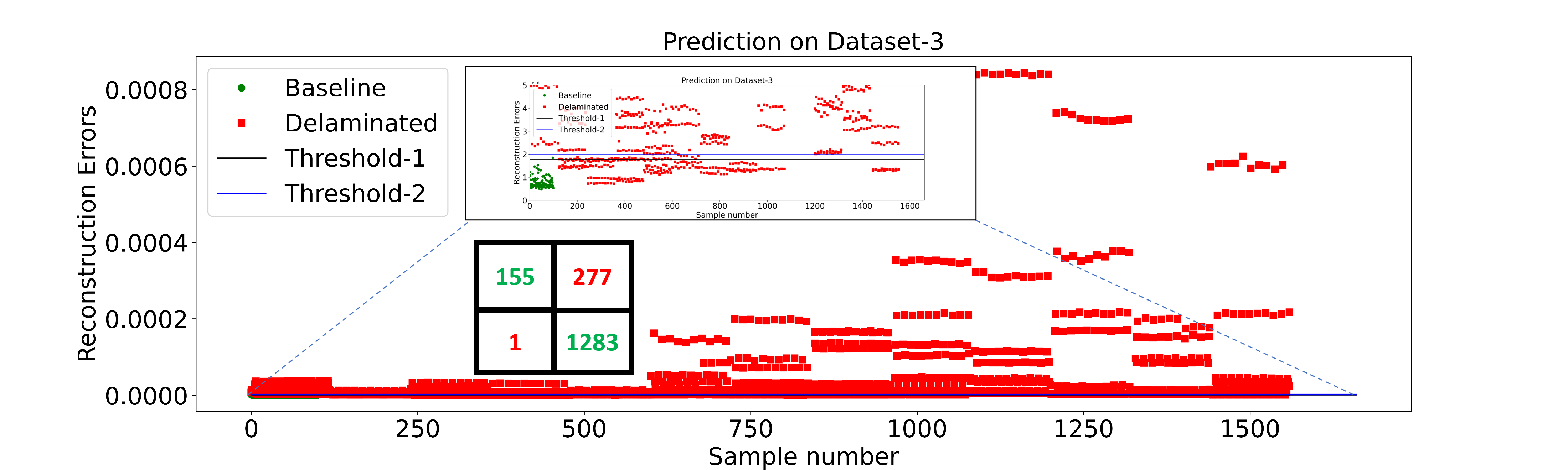}
	\end{minipage}
	\caption{CAE based delamination prediction on (a) Dataset-1, (b) Dataset-2, and (c) Dataset-3. The confusion matrix is represented by a 2$\times$2 box corresponding to each dataset.}
	\label{fig:threshold}
\end{figure}

After training the networks, we have followed the delamination predictions strategy as shown in Fig.~\ref{fig:strategy}. We have set two different thresholds, i.e., maximum value and 99$^{th}$ quartile of reconstruction errors of the training samples. If the MSE of the test sample is more than the threshold, then it is categorized as delamination. In Fig.~\ref{fig:threshold} we have plotted the reconstruction errors along with the confusion matrix of all test samples. The threshold-1 (black) is based on 99$^{th}$ quartile, whereas threshold-2 (blue) is the maximum value of reconstruction error. It is seen that for dataset-1 and 2, threshold-2 gives better prediction accuracy, whereas threshold-1 works better for dataset-3. One crucial point to recall here is that the network is trained only on baseline training examples and making predictions on new baseline and delamination samples. The networks has achieve high accuracy across three datasets, nearly 100\% for the first two and nearly 85\% for the third dataset. From Fig.~\ref{fig:threshold}(c), similar reconstruction errors is observed for baseline and misclassified damaged signals. These misclassifications for dataset-3 are majorly due to false positives, i.e., the delamination signals categorized as baseline signals. However, the type of misclassifications (false positives and false negatives) depend on the threshold settings. The shifting of threshold-1 downwards will increase the false negatives and reduce the false positives. It is observed that multiple signals in the dataset corresponding to delamination look similar to the baseline signals. The signal change is proportional to the interaction of the signal with the delamination, and an unfavorable position of a particular actuator-sensor pair may lead to a minor interaction between the signal and damage. Due to this, such damage signals will be classified as normal. Beside this, the dataset is designed for computed tomography applications, not specifically for deep learning implementations.

\begin{figure}[h!]
	\centering
	\begin{minipage}[b]{0.32\textwidth}
		\centering
		\includegraphics[width=1.0\textwidth]{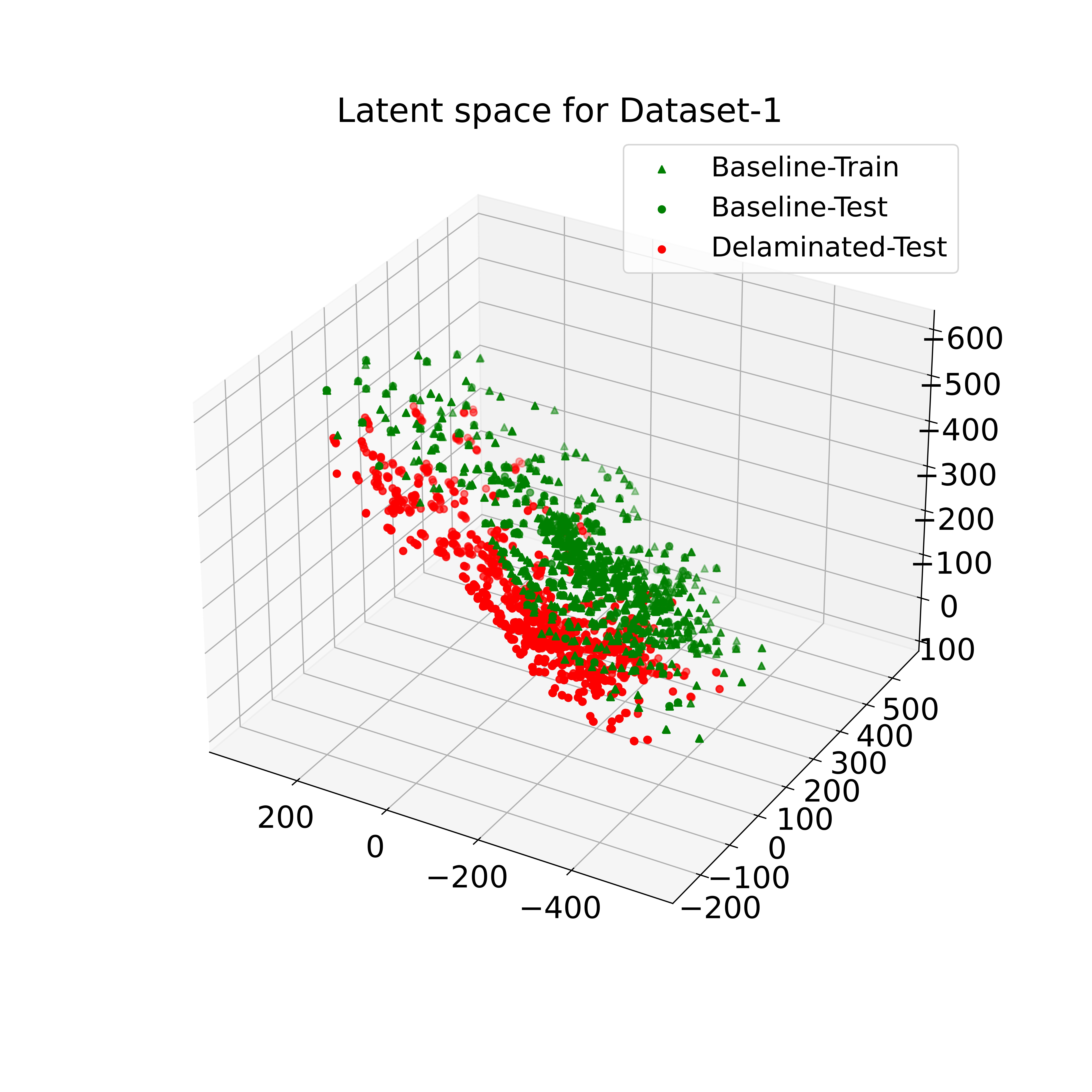}
	\end{minipage}
	\begin{minipage}[b]{0.32\textwidth}
		\centering
		\includegraphics[width=1.0\textwidth]{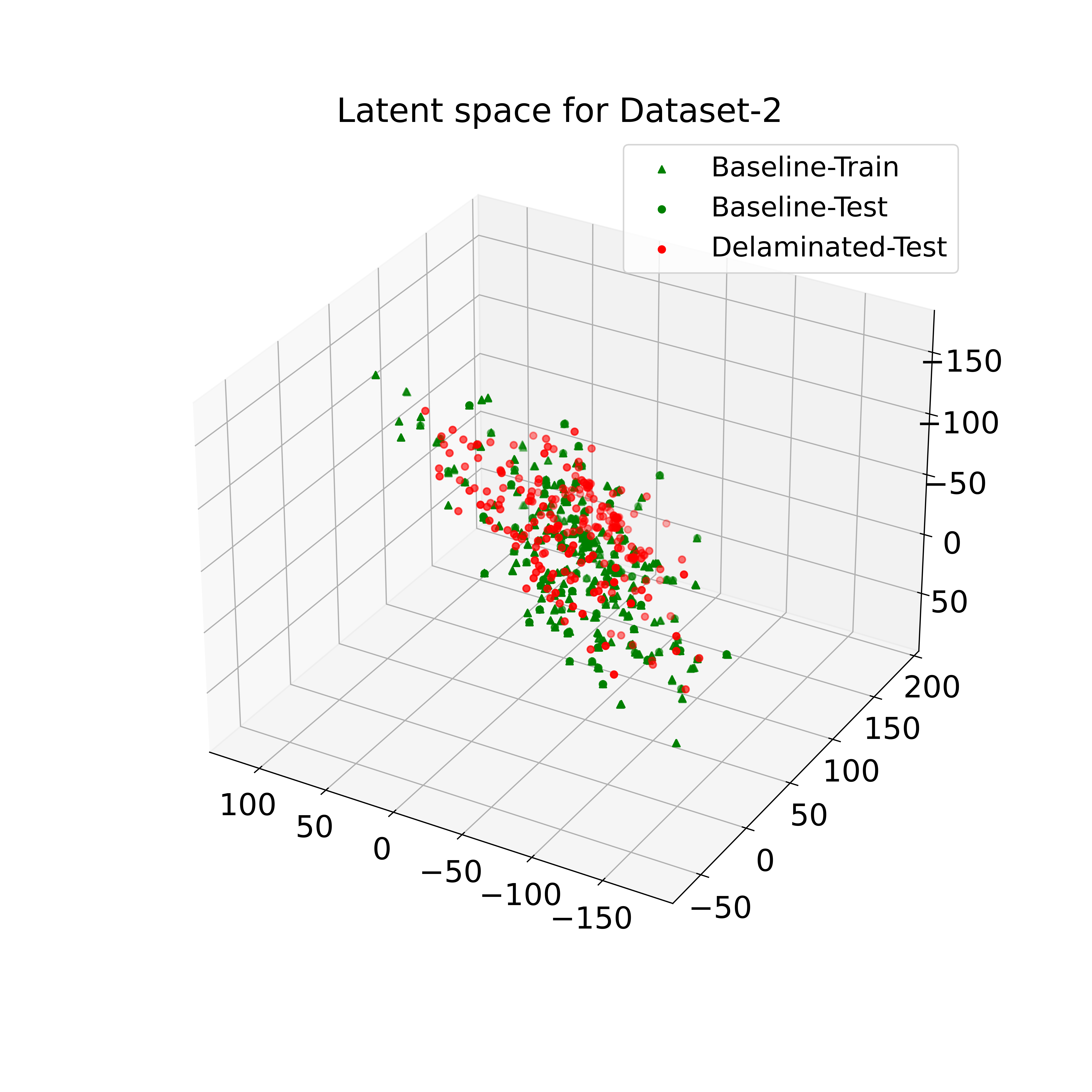}
	\end{minipage}
	\begin{minipage}[b]{0.32\textwidth}
		\centering
		\includegraphics[width=1.0\textwidth]{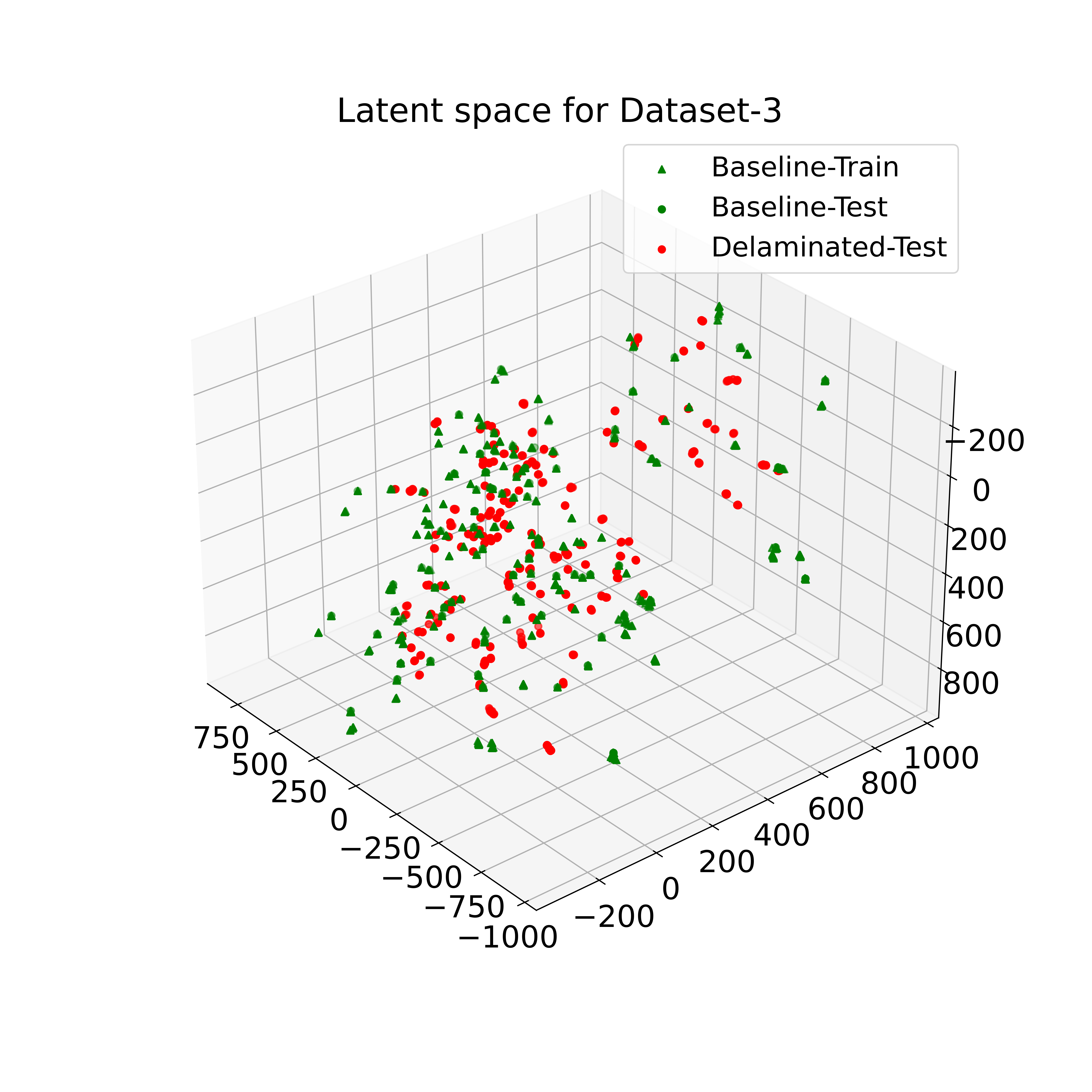}	
	\end{minipage}
	\caption{3D latent space visualization of CAE for three case studies. Green and red dots represent test signals corresponding to baseline and delamination respectively.}
	\label{fig:latent}
\end{figure}

We have plotted the three-dimensional latent space of the networks for all three datasets, as shown in Fig.~\ref{fig:latent}. It can be observed that the latent space of baseline and delaminated signals are segregated into two separate clusters for dataset-1 (leftmost in Fig.~\ref{fig:latent}). This distinction starts diminishing towards dataset-3 (rightmost in Fig.~\ref{fig:latent}). The reconstruction ability of CAE (decoder to be specific) depends upon the generalizations in the latent space. A well-clustered latent space gives better reconstructions than a scattered latent space. This can be confirmed from the average reconstruction loss, which is lowest for dataset-1 (2.5e-7) followed by other dataset-2 (5e-7) and dataset-3 (8e-7). Besides this, it can be seen by comparing Figs.~\ref{fig:threshold} and \ref{fig:latent} that better separation between baseline and delaminated samples in the latent space with lower variance (Fig.~\ref{fig:latent}) is reflected directly on the separation and the distribution of both types of samples across the threshold (Fig.~\ref{fig:threshold}).

We have trained PCA-ocSVM, ICA-ocSVM and CAE on a CPU configuration of i7- 9700KF (8 cores) with 32 GB RAM, GPU of Nvidia RTX-2070 (CUDA cores = 2304 and Tensor Cores = 288) with 8 GB of VRAM. The code files are developed in Python programming language and the code package is available on Github.

\section{Comparisons \& Discussions} \label{sec:results}

\begin{figure}[h!]
	\centering
	\includegraphics[width=1.0\textwidth]{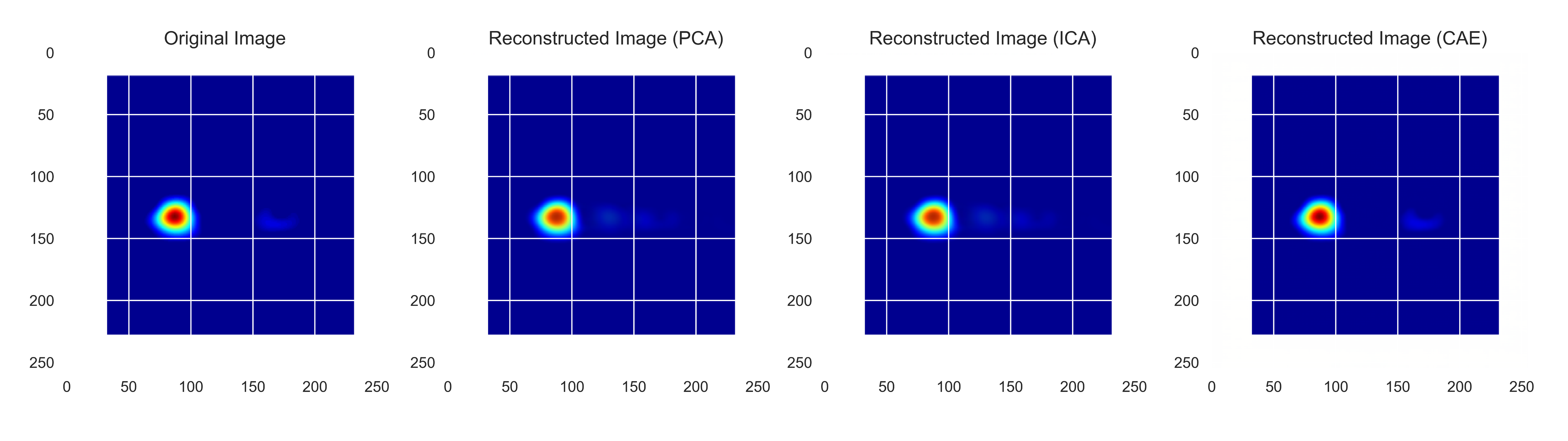}
	\caption{(a) Original image and reconstructed image from (b) PCA, (c) ICA and (d) CAE.}
	\label{fig:recon_all}
\end{figure}

In this study, we have implemented three methods, i.e., PCA-ocSVM, ICA-ocSVM, and CAE for delamination prediction. The first step is to construct a lower-dimensional latent space followed by relationship learning. For PCA and ICA, inverse transforms can be used to generate the reconstructions from the latent space. On the other hand, reconstructions are the output of the trained network in CAE. We have illustrated one sample from dataset-3 along with the corresponding reconstructions from PCA, ICA, and CAE in Fig.~\ref{fig:recon_all}. 

From the figure, it is seen that both PCA and ICA are able to reconstruct the main blob of the original image, whereas it is facing problems in reconstructing the smaller blobs. On the other hand, CAE is able to reconstruct the image perfectly. The reconstruction error between the original and the reconstructed image is calculated as 1e-4 for PCA and ICA and 1.9e-6 for CAE, respectively. For the dataset-3, the average reconstruction error of the training samples is measured as 2e-4 for both PCA and ICA, whereas 8e-7 for CAE. One reason for this difference lies in how PCA, ICA, and CAE perform the dimensionality reduction (or feature extraction) task. CAE uses non-linear functions to construct more powerful generalizations in the latent space than linear dimensionality reduction methods like PCA and ICA.

In the second step of relationship learning, ocSVM is used for ML-based methods, whereas CAE learns the relationship using the same network. Fig.~\ref{fig:accml} and \ref{fig:threshold} shows that CAE has provided higher accuracy on anomaly detection for all the three datasets. CAE is able to achieve 100\% accuracy on the first two datasets and above 85\% on the third dataset. However, the ML-based methods demand less computational time and resources than DL-based methods. 

With unsupervised feature-learning techniques, overfitting-generalization or safety-cost tradeoff plays an essential role in the success of anomaly detection tasks. In ML-based methods, the balance of false positives and false negatives is controlled via hyperparameter $\nu$, whereas CAE uses thresholds. However, handling thresholds are more convenient because it provides different safety-cost tradeoff without training the algorithms from scratch. 

\section{Conclusions} \label{sec:conclude}
In this paper, we have presented unsupervised-feature learning methods for delamination prediction in composite panels. We have used wavelet-enhanced representations for better featurization of the guided wave datasets. We have used three benchmarks experimental datasets that are different from each other in multiple aspects. We have trained two ML-based (PCA-ocSVM, ICA-ocSVM) and one DL-based (CAE) unsupervised feature learning algorithms on baseline training samples from the three datasets. We have utilized the trained algorithm to detect the delamination for new set of baseline and damaged signals. The entire anomaly detection methodology is based on two aspects: Feature extraction and relationship learning. In ML-based methods, we have used PCA and ICA-based linear dimensionality reduction techniques for feature extraction. Then, we have trained ocSVM on the extracted features to learn the distribution of baseline signals. On the other hand, both dimensionality reduction and relationship learning is performed under the same NN architecture in CAE. We have observed that CAE can learn better features and reconstruct the input space with less mean squared reconstruction error. We have seen that the accuracy of CAE for delamination detection exceeds the ML-based methods. CAE is able to achieve high accuracy on all the datasets.

\bibliography{mybibfile_cs}
\end{document}